# Numerical and Experimental Characterization of RF Waves Propagation in Ion Sources Magnetoplasmas

G. Torrisi, D. Mascali, G. Sorbello, G. Castro, L. Celona, *Member, IEEE*, and S. Gammino

*Abstract*—This paper describes three-dimensional numerical simulations and Radio Frequency (RF) measurements of wave propagation in microwave-heated magnetized plasmas of ion sources. Full-wave solution of Maxwell's equations has been addressed through the Finite Element Method (FEM) commercial software COMSOL. Our numerical model takes into account the strongly inhomogeneous and anisotropic magnetized "cold" plasma medium. The simulations reproduce the main features of the wave-plasma interaction of the Flexible Plasma Trap (FPT) that recently came into operations at INFN-LNS. A two-pins RF probe has been ad-hoc developed and used as a plasma-immersed antenna for measuring local wave electric fields in the FPT device. The measurements of plasma electron density and RF electric field, performed for different external magnetic-field configuration, allowed a direct comparison with the assumed simulation model.

## I. Introduction and Motivation

RADIOFREQUENCY propagation of electromagnetic waves through plasmas [1], [2], has gained considerable interest in many fields: broadcast communication space exploration [3], plasma sheath in reentry vehicles [4], thermonuclear fusion [5], plasma processing [6], microwave plasma diagnostics and heating in ion sources [7]–[10]. In particular, we focus on the mechanism of RF propagation into the non-homogeneous magnetized anisotropic lossy plasma of Electron Cyclotron Resonance Ion Sources (ECRIs) and Microwave Discharge Ion Sources (MDIS) [11]–[13].

The issue of radiofrequency propagation in anisotropic plasmas confined in compact traps such as ion sources is still a hot-topic in the relative scientific community. No models and results from numerical approaches exist in literature providing a clear and reliable description of wave launching, propagation and absorption in the small-size (compared to the wavelength) magneto-plasma. The point of a direct comparison to experimental data, in particular to spatially resolved measurements of the in-plasma electric field, is also of particular importance, since it could represent a step towards the development of numerical tools becoming more and more predictive.

In this paper, the electromagnetic wave propagation in microwave-heated plasma confined in a magnetic field, is addressed through numerical simulations and experiments, devoted to understand how the RF wave propagation depends on the electron density profile and external magnetic configuration which can be often controlled in many kind of plasma physics and application experiments.

A better comprehension of electromagnetic wave propagation and RF power to plasma coupling mechanisms in ion-sources magnetoplasmas is crucial in order to provide a cost-effective upgrade of these machines alternative to the use of higher confinement magnetic fields, higher RF power level and higher pumping wave frequency [14].

Three-dimensional self-consistent numerical simulation of full-wave RF fields in inhomogeneous non-uniformly magnetized plasma is a big challenge [15], [16] and only few methods have been developed to obtain self-consistent solutions [17], [18]. Several full-wave numerical simulation codes have been developed, exploiting ray-tracing technique or finite-difference time-domain (FDTD) method [19]–[22]. Alternative models and numerical strategies have been attempted by other authors by assuming assuming "zero-dimensional" up to 2D-simulation model in a simplified magnetostatic scenario with an axial symmetry or by considering the plasma medium as an equivalent dielectric load [23], [24]. However, the wave propagation at frequencies in the range 3-40 GHz in ECRIS compact plasma (having magnetic field of the order of few Tesla, electron density in the order of $10^{18}$ m$^{-3}$) cannot be predicted by the plane wave model nor addressed by "ray-tracing". These models, in fact, fail in minimum-B configuration scenarios where the scale length of plasma nonuniformity, $L_n = |n_e/\nabla n_e|$, and the magnetostatic field nonuniformity, $L_{B_0} = |B_0/\nabla B_0|$, are smaller than the free space, $\lambda_0$, and guided, $\lambda_g$, wavelengths. The Wentzel, Kramers and Brillouin (WKB) method - often adopted to describe waves in fusion toroidal devices [25]–[27] - is also not applicable because the wavelength is comparable with the plasma gradient and dielectric tensor scale lengths. Only full-wave methods are suitable to capture waves' propagation features, strongly affected by the resonant cavity leading to the formation of electromagnetic standing wave modes.

Hereinafter we present numerical results of 3D simulations of RF wave propagation in the magnetized plasma of the "Flexible Plasma Trap" (FPT) [28], an ion source test-bench developed on purpose at INFN-LNS, to testing novel RF systems for plasma heating and diagnostic schemes. We used COMSOL Multiphysics [29] software to model a "cold", anisotropic magnetized plasma, described by full-3D non uniform dielectric tensor, enclosed by the metallic cylindrical cavity where the plasma is generated. A proper mesh generation, exploiting FEM-based COMSOL versatility, allowed us

The authors wish to thank the 5th Nat. Comm. of INFN, under the Grant PANDORA for the financial support.
G. Torrisi (e-mail: giuseppe.torrisi@lns.infn.it), D. Mascali, G. Castro, L. Celona, S. Gammino are with the Istituto Nazionale di Fisica Nucleare - Laboratori Nazionali del Sud, Via S. Sofia 62, 95123 Catania, Italia
G. Sorbello is with Università degli Studi di Catania, Viale Andrea Doria 6, 95125, Catania, Italia and also with the Istituto Nazionale di Fisica Nucleare - Laboratori Nazionali del Sud, Via S. Sofia 62, 95123 Catania, Italia to optimally modelize the cavity and microwave waveguide launching structure, with a good computational efficiency and high resolution especially around the resonance regions. The main goal of the paper is to directly compare, for the first time in ECR ion sources, the numerical models results with RF measurements of the wave amplitude inside the FPT plasma chamber, performed through a two-pin RF probe antenna.

Measurements of field structure correlated with plasma density, magnetic profile and electron temperature have been already presented up to now only for radially localized helicon (RLH) waves detection on helicon plasma sources [30]–[32] and for direct detection of Electron Bernstein Waves (EBW) in large volume magnetoplasmas of toroidal fusion plasma reactor [33]–[35]. In the latter case the pumping wavelength $\lambda \ll L_n, L_{B_0}$. On the other hand, in ECRIS complex environment, the strong interaction of the resonant cavity with the electromagnetic wave propagation affects the polarization state and no a priori assumption on the polarization state can be done. RF measurements in small volume magnetoplasmas are especially needed in order to investigate well beyond the assumed oversimplified damping mechanism of the right-hand polarized wave (RHPW) absorption at the ECR layers.

The article is arranged as follows. Section II is on wave field solutions of the Maxwell's equations taking into account the magnetic field which makes plasma anisotropic, non-uniformity of plasma density, and the metallic plasma chamber. The section III describes the experimental setup. Finally the results of the numerical analysis are presented in Sec. IV and compared with experiments.

## II. RF Wave-plasma interaction modeling

In compact ECR plasma machines, the magnetically confined plasma is ignited by microwaves in the GHz range resonantly interacting with free-electrons in a rarefied gas (down to $10^{-8}$ mbar) via Electron Cyclotron Resonance, occurring when $\omega_{RF} = \frac{eB_{ECR}}{m_e}$, where $m_e$ and $e$ are the electron mass and charge, respectively.

A magnetized plasma in the GHz range frequencies can be modeled as a cold magneto-fluid with collisions where the field-plasma interaction is described by the tensorial constitutive relation $\overline{\overline{\varrho}} \cdot E$. Tipically $\overline{\overline{\varrho}}$ is derived assuming a magneto-static field $B_0$ directed along just one axis. This assumption is valid in most of cases but not in ECRIS where $B_0$ is not strictly axis-symmetric. Considering the actual magneto-static structure of an ECRIS, that is not uniform nor axis-symmetric, $\overline{\overline{\varrho}}$ depends in a complex way from the magnetostatic field $B_0(x, y, z)$ and the local electron density $n_e(x, y, z)$. Under the "cold plasma" approximation, (i.e. $v_\varphi \gg v_{th}$, being $v_\varphi$ the wave's phase speed and $v_{th}$ the electron thermal speed), the tensor components have been calculated in [13] where it is shown that the tensor $\overline{\overline{\varrho}}(x, y, z)$ exhibits a local dependence from the external static magnetic field, $B_0(x, y, z)$, and plasma density, $n_e(x, y, z)$:

$$\overline{\overline{\varrho_r}}(x,y,z) \Rightarrow \overline{\overline{\varrho_r}}(B_0, n_e) \qquad (1)$$

Eliminating the magnetic field between Maxwell's equations and using the above constitutive relation (1) for an anisotropic medium, the wave equation reads as:

$$\nabla \times \nabla \times E - \frac{\omega^2}{c^2}\overline{\overline{\varrho_r}} \cdot E = 0 \qquad (2)$$

The above wave equation (2) can be solved as a driven problem by a FEM solver that supports a non homogeneous tensorial constitutive relation; in the present work we used COMSOL and an external MATLAB routine allowing the definition of the full 3D dielectric tensor. For sake of clarity, the simulations hereby described are based on the "stationary" frequency domain COMSOL solver since general properties of an ECR plasma can be considered stationary on very long timescales.

In COMSOL we were able to model the exact shape of the FPT plasma chamber and the waveguide launching structure.

While the static magnetic field, $B_0$, is externally imposed by the current flowing in the magnetic system and can be both accurately predicted and measured (and for this reason can be considered a known input parameter), on the contrary, the plasma density, $n_e$, depends from the local electromagnetic field, $\{E, H\}$, that sustain the plasma. In other words the anisotropic wave equation (2) and constitutive relations (1) are in principle a system of coupled equations. The coupled problem can be solved by an open loop approach [13] assuming a reasonable plasma density and or in a self-consistent approach [18].

In this work we overcome the above approaches and we adopt the empirically meeasured magnetic field and the plasma density. Then, for the first time, we compare the numerically predicted electromagnetic field, $\{E, H\}$, with the measured one. The validation has been carried out on a particular machine (the FPT), however it should be noted that the approach remains of very general interest since the constitutive relation are local (1) and valid for any anisotropic cold plasma. Moreover, the FPT allows a wide range of working points that have been explored in the experimental validation carried out (see below). Since in proximity of the resonance surface (individuated by the iso-surface $B_0 = B_{ECR} = \frac{m e}{e \omega}$) the permittivity varies strongly, the discretization of such a narrow region needed a very fine mesh: to achieve this, we adopted an adaptive mesh procedure (that was allowed by a specific feature of the solver). An extremely fine mesh has been obtained, thanks to "functional evaluation" based on the electric field gradient. In the simulations, the experimental setup is modeled by the computational domain shown in Fig. 1. The domain consists of the cylindrical plasma chamber cavity and microwave double ridged injection waveguide operating in the fundamental $TE_{10}$-like mode to allow a wide bandwidth (from 3.5 GHz to 8.20 GHz). The inner cavity volume is filled by lossy plasma characterized by dielectric tensor. The cavity walls are modeled via the appropriate "Perfect Electric conductor" boundary condition. Table I reports the Simulation input parameters.

## III. Experimental Scenario and RF probe antenna description

The Flexible Plasma Trap (FPT) installed at INFN-LNS and described in [28] consists of a copper cylindrical plasma chamber 250 mm long and 65 mm in radius, having numerous flanges for pumping, gas inlet, diagnostic and three RF rectangular input waveguide ports for test different plasma heating schemes. Figure 3 shows a schematic diagram of the FPT machine, the RF power delivery system, the three magnetic coils, and diagnostic probes (RF probe for electric field detection or Langmuir Probe for electron density measurement). The FPT plasma has been fed by continuous mode microwaves at the operating frequency of 6.827 GHz generated by a TWT Microwave Amplifier. An isolator protects the TWT from any reflected power and a dual directional coupler is used for measuring the forward and reflected powers. Microwaves are then guided via WRD350 double ridge rectangular waveguides and through a Kapton window into the vacuum chamber which maintains the vacuum integrity. Plasma is confined by a static magnetic field obtained by means of three coils, placed coaxially to the plasma chamber and able to generate different magnetic field profiles. As it is possible to see in Fig. 3, one of the chamber endplates is dedicated to microwave launching and gas-injection, the other one to plasma diagnostics. The experiments are performed at wave input power $P_{in}=100$ W at the operating frequency of 6.827 GHz and at the pressure of $1.5 \cdot 10^{-4}$ mbar. We used three different configuration for the external magnetic field leading to three "simple mirror"-like profiles.

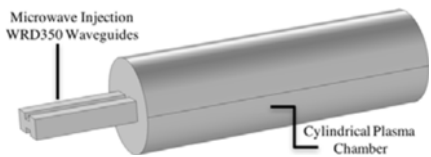

Fig. 1: Simulated Geometry: Cylindrical plasma chamber cavity and microwave rectangular double ridge waveguide injection

TABLE I: Simulation input parameters.

| Parameter | Value | Description |
|---|---|---|
| L | 250 [mm] | Cavity length |
| R | 65 [mm] | Cavity radius |
| $\nu_{RF}$ | 6.827 [GHz] | Frequency |
| $\omega_{eff}$ | $\omega_{cp}/10^3$ [rad/s] | Collision frequency |
| $P_{RF}$ | 100 [W] | RF Power |

A ceramic-coated RF two-pins probe small antenna, with one pin grounded and another connected to the inner wire of a coaxial transmission line, has been developed and used to measure the wave electric field inside FPT plasma chamber. In principle, we could use a simple semi rigid coaxial cable with a coaxial pin that protrudes from the dielectric insulation (low density dielectric Teflon® PTFE). However, a two-pins probe antenna allows to measure the potential difference (or the E-field component) perpendicular to the pins resulting in polarization selectivity as demostrated in [35].

The pins length and distance have been designed to be sensitive to the short wavelengths and small enough to achieve the desired spatial resolution. The probe is constructed out of Sucoflex© microwave 0-40 GHz cable, having a silver-plated copper inner conductor with diameter of 1 mm and an outer diameter 4 mm. The probe exposes two linear tips having length of 3.5 mm and a distance of 2 mm, and it is enclosed in a high-purity alumina protection tube allocated parallel to the longitudinal axis plasma chamber. The probe is connected to vacuum RF feedthrough K-connector that was mounted on a standard CF-40 flange. A bellow allows movements via computer controlled step motor which provides a 0.1 mm precision. In the current set up (see Fig. 3), the probe can penetrate into the plasma chamber along a line parallel to the chamber axis. The RF probe pins are protected from direct contact with the plasma by a coating of MACOR. The output from the RF probe is fed into Power Probe. The new electric field probe has many desirable features such as high sensitivity, good spatial resolution, polarization sensitivity, and proven reliability in a high temperature environment. Together with the already existing diagnostics for X-ray, optical emission, electron density, it gives the possibility to characterize plasma also in the RF electromagnetic spectrum.

IV. NUMERICAL AND EXPERIMENTAL RESULTS

Full-wave simulations were carried out through the COMSOL Multiphysics FEM solver, considering the geometry shown in Fig. 1 and the above described "cold plasma" model. Our three-dimensional 3D RF field solver uses the experimentally measured plasma density, $n_e(x, y, z)$, and magnetostatic, $B_0(x, y, z)$, profiles as input functions. Figure 2 displays three different cases of electron density $n_e$ measured through a Langmuir probe [14] and obtained for three different values of the ratio $\frac{B_{min}}{B_{ECR}}$ between the minimum magnetic field, $B_{min}$, and the value of magnetic field on the resonance surface, $B_{ECR}$.

In particular, concerning the plasma density measurements, they have been obtained along a line of view that is almost parallel to the plasma chamber axis. Regarding the radial profiles, at the moment we do not have direct access for measuring the density along a given line of sight. Anyway, in simple mirror devices, considering basic assumptions about plasma diffusion across the magnetic field, a $\cos(r)$ trend can be reasonably argued and assumed for radial density trend as suggested in (36), and it is in general agreement with the optical inspection of the plasma and with X-ray imaging (37).

The magnetic field map has been experimentally obtained measuring the magnetic induction by means of a commercial "Lake Shore Cryotronics, Inc." Hall probe. The field has been measured in axis and off-axis position, both for axial and radial components, then the 3D map can be easily reconstructed taking profit from the symmetry of the system. In particular, we study three working points differing for the magnetic field configurations:

1) Simple mirror configuration with a z-distance between ECR layers $W_{ECR}$ (where $B_0 = B_{ECR}$) of 70 mm and a ratio of the minimum magnetic field $B_{min}$ over the $B_{ECR}$, $\frac{B_{min}}{B_{ECR}} = 0.65$.

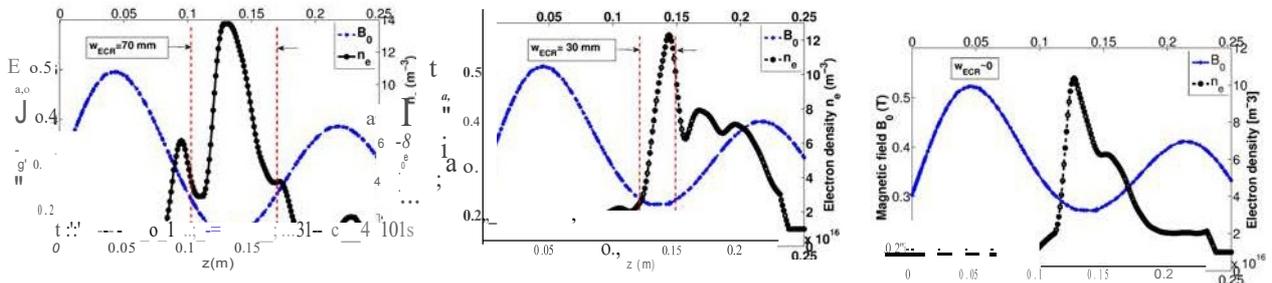

(a) Case I $(w_{ECR} = 70$ mm$)$, Input parameters: black point marker curve represents plasma density profile and blue plus marker curve static magnetic field $B_0$ for $\frac{\omega}{\omega_{ECR}} = 0.65$.

(b) Case 2 $(w_{ECR} = 30$ mm$)$, Input parameters: black point marker curve represents plasma density profile and blue plus marker curve static magnetic field $B_0$ for $\frac{\omega}{\omega_{ECR}} = n$.

(c) Case 3 $(w_{ECR} = 0$ mm$)$, input parameters: black point marker curve represents plasma density profile and blue plus marker curve static magnetic field $B_0$ for $\frac{\omega}{\omega_{ECR}} = 1$.

Fig. 2: Measured electron density and magnetic profile along the the z-axis; this data is used as input parameter for the simulations, the positions of the ECR layers are indicated with vertical red dashed lines.

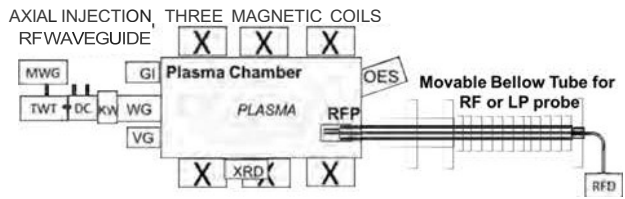

Fig. 3: Schematic of the FPT experimental setup at the INFN-LNS. MWG: 1-20 GHz Microwave Generator. TWT: 500 Watt 4-7 GHz Traveling Wave Tube Amplifier, DC: directional coupler, KW: Kapton window, WG: waveguide section, VG: Vacuum Gauge, GI: Gas Injection, RFP: Radio Frequency Probe, LP: Langmuir probe, XRD: X-Ray detector, OES: Optical Emission Spectroscopy, RFD: Radio Frequency detector DC-44 GHz.

2) Simple Mirror configuration with $w_{ECR} = 30$ mm, $\frac{\omega}{\omega_{ECR}} = 0.9?$
3) Simple Mirror configuration without ECR layers and $w_{ECR} = 0$ mm, $\frac{\omega}{\omega_{ECR}} = 1$.

For each working point, the "cold" permittivity tensor for 3D full-wave simulations was derived and used.

In order to explore the different propagation characteristics in terms of dissipated power (RF power absorbed by plasma) we can see the simulation results in figures 4b and 4c. The dissipated power - plotted on xz plane and along the line of sight of the RF probe - allows a comparison of the power absorption for the three magnetic field configurations listed above. Both ID and 2D plots show that the power is mostly deposited along layers, which become closer and closer when increasing the $\frac{\omega}{\omega_{ECR}}$ ratio. 2D plots point out the structure of the absorption layers which is strongly modified by the ratio as well. In particular, the model is able to capture the physics involved, since absorption clearly occurs where ECR layers are placed at each ratio, as it is clearly shown in Fig. 4a.

The direct comparison between simulations and measurements demonstrate that the COMSOL simulated RF field profile clearly resembles the experimentally measured ones (see Fig. 5) for the three magnetic fields configurations showed in Fig. 2. Figures 5 show good agreement between the measured and full-wave FEM simulated field intensity $|E|^2$. What it is relevant is that for the configurations at larger $w_{ECR}$, the electromagnetic field seems to self-confine in the plasma volume enclosed by the resonance layers. This "plasma-cavity" effect was already predicted in [13], [38] occurring under specific conditions of plasma density and sizes. It is now confirmed also by the direct RF measurements. The amplitude peaks near the ECR layers confirm that the presence and position of the ECR layers can be predicted by the FEM calculations. It is however remarkable that - regardless from the absolute height of the single peaks - the simulations very well reproduce the peaks-valleys sequence (which is a direct consequence of wavelengths scale in the plasma filled cavity). It appears very clear that the agreement between experimental and numerical data is better when the plasma size is the largest one, while for the $w_{ECR} = 0$ case the profiles tend to deviate each other. In particular measurements show that the electromagnetic energy starts to "escape" from the inner resonance volume as $w_{ECR}$ becomes smaller and smaller. This effect is only partially reproduced by the simulations. It is worth mentioning, anyway, that the configuration $w_{ECR} = 0$ is very rarely adopted in plasma traps and was here mentioned as a critical condition for checking the model reliability under different magnetic field profiles. The agreement is higher as higher it is the plasma density within plasma chamber. This is probably caused by the perturbing presence of the radio frequency probe.

Up to now, simulations results are without probe; we know the perturbation introduced from the probe on the electromagnetic field strongly depends on the probe position. This in principle implies that we should perform one simulation for each probe position, which will be done in future works. For the moment we performed one additional simulation by considering also the presence of the probe for the case 2, $(w_{ECR} = 30$ mm, $\frac{\omega}{\omega_{ECR}} = 0.92)$ at a fixed position $z = 0.2$ m. in Fig. 6, we compared the simulated electric field with and without probe. Due to the presence of the insulating dielectric

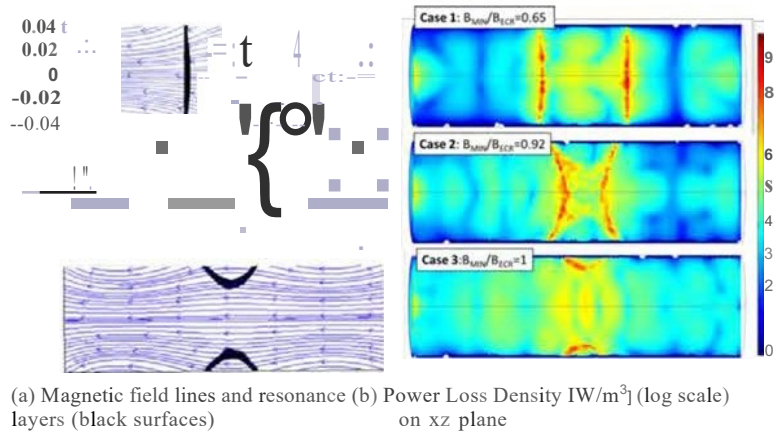

(a) Magnetic field lines and resonance layers (black surfaces)  
(b) Power Loss Density [W/m$^3$] (log scale) on xz plane

(c) Power Loss Density [W/m$^3$] (log scale) along the line of sight of the RF probe

Fig. 4: Simulated Electromagnetic Power Loss Density [W/m$^3$] (log scale) in plasma chamber.

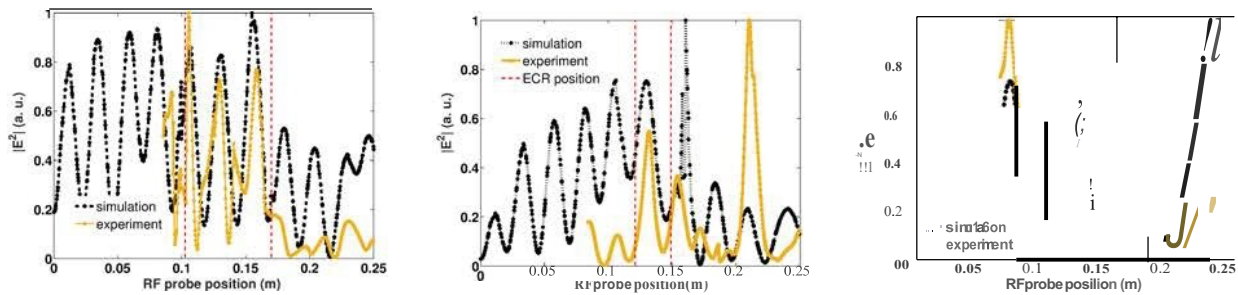

(a) Case 1: $w_{een}$ = 70 mm. $= 0.65$. This is the case with the largest plasma volume enclosed by the ECR layers. There is an excellent agreement between simulated and measured RF field distribution.

(b) Case 2: $w_{sen}$ = 30 mm, $= 0.92$. With respect to Case 1, the plasma volume is becoming smaller. The agreement between simulated and measured RF field distribution is still evident.

(c) Case 3: $w_{een}$ = 0, $= 1$. This represents a critical configuration, very rarely tuned in ECR machines. In this case there is still a general agreement even if locally the two plots deviate somewhere.

Fig. 5: Simulated and measured RF electromagnetic field along the probe position.

of the coaxial probe, an attenuation of the field can be noticed as it is evident also in the experimental results of Fig. 5 (b). For lower electron densities (case 3), the probe influence on the cavity mode dominates. When density increases, the modal behavior becomes more and more sensitive to the plasma presence within the chamber, whose influence is predicted by the FEM simulations. Therefore, the agreement between FEM predictions and experimental measurements increases (case 1).

## V. CONCLUSION

The paper has reported about the direct comparison of numerical results (from a full-wave approach) with the inner-plasma measurements of the electric field amplitude performed by means of a two-pins RF probe. Both simulations and measurements have been done considering a compact plasma trap (resembling very closely the features of any ECR ion source) in a simple mirror configuration. The resulting code that propagate the electromagnetic fields into a cold anisotropic plasma and its validation against experimental data is very accurate and of general effectiveness: we showed that the cold plasma model (1) captures the essential features of the wave propagation in plasma when both the static magnetic field and the plasma density are accurately estimated and/or tuned on experimental data. The excellent agreement between model predictions and experimental data are very promising for the design of future launchers or "exotic" shapes of the plasma chambers in compact machines, such as ECR Ion Sources and other similar devices. Our approach - integrating a FEM commercial solver and a self-made MATLAB developed routine for local tensor computation - starts to shed some light on the electromagnetic problem of propagation and absorption in magnetized plasma. Hence the results, even if benchmarked on a particular machine, have actually a general validity, at least for every RF-heated magneto-plasma compact device or wave propagation in magnetized plasma, in cold approximation. Further steps forward are going to be done as concerning the improvement of the model, including the "hot" plasma tensor: this will perspectively allow to master additional mechanisms such as modal conversion at the hybrid resonances.

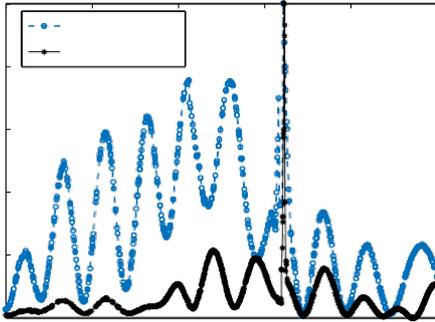

Fig. 6: Simulated Elecric field for Case 2 with and without probe.

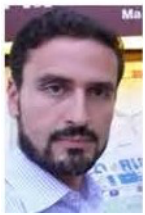

**David Mascali** David Mascali was born in Catania, Italy, in July 1981. He received the Master degree (in 2005) and the PhD (in 2009) cum laude in Experimental Physics from the Universil degli Studi di Catania, Italy. He is currently a Staff Researcher at the Laboratori Nazionali del Sud, Istituto Nazionale di Fisica Nucleare (CNFN-LNS), leading the Ion Beams Production team inside the Accelerators Division. He is the spokesperson of the PANDORA experiment, supported by INPN, aiming at interdisciplinary research in Astrophysics and Nuclear Physics in laboratory plasmas, and of the Grant "IRIS" for innovative resonators for ECR ion sources. His research activity is mainly focused on plasma physics, plasma-based ion sources science and technology, plasma diagnostics, laser produced plasmas. In 2009 he received, in Firenze, the National F. Resmini Prize for innovative research in Accelerator Physics. In 2012, he was awarded in Sidney, Australia, by the international "R. Geller" prize for outstanding researches in Ion Sources Science and Technology. He is a member of the Scientific Programme Committee of Int. Conf. on Ion Sources since 2013. Since 2017 he teaches classes of "Plasma Physics for Interdisciplinary Research" at the PhD course of the Catania University. He gave several invited talks and seminars in Europe, USA and Japan, authoring more than 80 peer-reviewed papers and more than 70 conference proceedings.

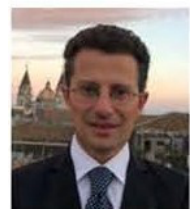

**Gino Sorbello** Gino Sorbello received the degree in Electronics Engineering cum laude at the University of Catania, Italy, in 1996, and the Ph.D. degree in Electronics and Communications Engineering at the Politecnico di Milano in 2000. In 2002 he became Assistant Professor of Electromagnetic Fields al University of Catania. Since 2014, he is Associate Professor of Electromagnetic fields at the Dipartimento di Ingegneria Elettrica, Elettronica e Informatica, University of Catania. His current research interests include the study of single-mode solid-state waveguide lasers and amplifiers, integrated optics, the development of planar antennas and ultra-wideband compact antennas and antenna arrays, the study of microwave devices and computational electromagnetism with a special interest in RF-plasma interactions and particle accelerators. Gino Sorbello authored more than 50 scientific papers and more than 30 conference proceedings.

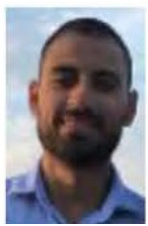

**Giuseppe Torrisi** Giuseppe Torrisi was born in Catania, Italy, in 1987. He received the M.S. Laurea degree (summa cum laude) in telecommunication engineering from the University of Catania (Catania, Italy) in 2011 and the Ph.D degree in information engineering from Universita Mediterranea di Reggio Calabria (Reggio Calabria, Italy) in 2016. He is currently a temporary staff researcher of the Istituto Nazionale di Fisica Nucleare (INFN) at Laboratori Nazionali del Sud (LNS), Catania. His scientific activity is concerned with electromagnetic propagation in microwave generated plasmas produced by ion sources for particle accelerawrs. In particular, he is presently focused on the development of innovative microwave schemes for plasma heating and diagnostics. He is author of more than 50 papers published on peer reviewed scientific journals or international conference proceeding. Dr. Torrisi was the recipient of the Best paper prize during the XXI Riunione Nazionale di Elettromagnetismo in 2016, the France-5coResmini award for the best PhD Thesis on accelerators and new technologies from the INFN Comm. V in 2017, the Young Scientist Award at the 2nd URSI Atlantic Radio Science Meeting and the Sannino prize awarded by the Italian Electromagnetics Society for the best work on mm-waves and microwave circuits in 2018.

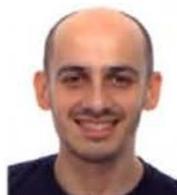

**Giuseppe Castro** Giuseppe Castro received the master degree in physics at the University of Catania, Italy, in 2008, and the Ph.D. degree in physics in 2013 in the same university with a thesis on the Study of innovative plasma heating methods and applications to high current ion sources. From 2015 he is adjunct professor of Institutes of physics I at the department of physics of the University of Catania. He is involved from the beginning of his career in the study of plasma physics devoted to application in the field of nuclear physics and ion sources, such as the development of innovative methods of plasma heating and plasma diagnostics. Among them, it is worth to cite the development of the optical emission spectroscopy, Langmuir probe, X-ray diagnostics and interfero-polarimetry for the magnetized plasma of ion source, studies on the electron Bernstein waves heating and the development, along the years, of different proton and ion source for accelerator, as VIS, PS-ESS and AISHa. In these fields, he authored more than 100 between peer-reviewed and proceedings and gave several talks in the most important topic conferences.

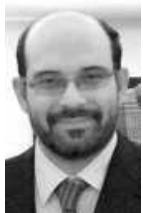 Luigi Celona L.G. Celona received the degree in Electronic Engineering at the University of Catania, in 1995 and he joined the Istituto Nazionale di Fisica Nucleare in 1996, at the Laboratori Nazionali del Sud (INFN-LNS), becoming Technological Engineer (Tecnologo) in 1998 and Principal Technological Engineer (Primo Tecnologo) in 2006. His main field of activity covers all the aspects of the production of singly and highly charged beams together with their acceleration to increase the performances of Particles Accelerators for Nuclear and Applied Physics. Experienced in all the design stages of an ion source: from mechanical design and manufacturing through the installation and final commissioning, he is also active in research and development, proposing different innovative concepts concerning the role of microwaves in the development of ECR and microwave ion sources. He is a member of the INFN Machine Advisory Committee as ion source expert and of the Conseil Scientifique et Technique du Dpartement des Acclrateurs, de Cryognie et de Magntisme of CEA. He was member of the steering committee of the SPES project and, in the European framework, he was a referee committee member of ARES and EMILIE projects to coordinate the R&D activities on ECR ion sources of the major European physics labs. He was the technical and scientific responsible of a joint-venture between INFN-LNS and some SMEs to design, realize and test a new hybrid ion source for Hadrontherapy named AISHa: two ion sources have been manufactured and successfully commissioned. He is the Leader of the design, manufacturing and commissioning of the high intensity proton sources along with the low energy beam transfer line for the European Spallation Source (ESS). The outstanding commissioning results fully comply the requirements given. The first source has been successfully installed in its final position at ESS site at the beginning of 2018 as a first part of ESS linac. He has also designed and built other types of ion and plasma sources, as the microwave discharge ion sources named MIDAS, TRIPS and VIS, for high efficiency ionization of the reaction products originating from an ISOL target and for intense monocharged production of light ions. During the period 2004-2007 he focused his efforts on the EXCYT radioactive beam facility, coordinating the installation, the commissioning and permitting to deliver the 8Li beam for the first experiments. He also worked on the development of the K-800 Superconducting Cyclotron bunching system contributing to the optimization of the cyclotron in the years 1995-1998 and to the axial injection beamline. He was the chair of the last International Workshop on ECR ion sources authoring more than 100 peer-review journals and more than 150 conference proceedings.

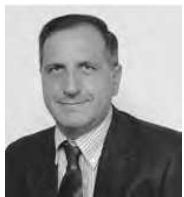 Santo Gammino Santo Gammino was born in Riposto, Italy, in 1963. He received the "Laurea" degree in Physics from the University of Catania, Italy in 1987 and he joined the Istituto Nazionale di Fisica Nucleare in 1988, at the Laboratori Nazionali del Sud in Catania (INFN-LNS), becoming Research Physicist in 1990, Principal Research Physicist in 2002 and Director of Research in 2009. His research has been focused to the production of monocharged and highly charged beams, and their acceleration. He proposed different innovative concepts for the development of ECR (electron cyclotron resonance) ion sources, and he actively worked on the construction of many different ECR ion sources. In particular his project SERSE came in operation in 1997 and for many years it was on the forefront for the production of highly charged ions. This activity has taken in the last years to the development of the so-called "3rd generation ECRIS and of different ion source (e.g. AISHA). He has also designed and built other types of ion sources, as laser ion sources and the microwave discharge ion sources named MIDAS, TRIPS and VIS, for high efficiency ionization of recoils and for intense monocharged beam production respectively. He also worked during the Nineties on the development of the K-800 Superconducting Cyclotron and on the design and construction of the EXCYT radioactive beam facility at INFN-LNS. Since 2010 he has been leader of the WP3-Normal conducting Linac for the European Spallation Source and for this facility he designed the high intensity proton source PS-ESS. He is also responsible for INFN in-kind contribution, including the Drift Tube Linac and the Medium Beta Superconducting Cavities, both under construction. Dr. Gammino served as a member of the National Committee of INFN for the Technological Research from 1996 to 2002 and for a third term from 2008 to 2012. In 2012 he has been charged as responsible for the R&D of Accelerators at INFN-LNS. He has been involved in the preparation of the NUPECC Long Range Plan 2010 and has been member of different Committees in Italy and abroad. Dr. Gammino has been author of more than two hundreds peer reviewed papers and more than three hundreds contributions to conferences and other publications.